# Head-to-Nerve analysis of electro-mechanical impairments of diffuse axonal injury

I. Cinelli[1], M. Destrade[2], P. McHugh[1], A. Trotta[3], M. Gilchrist[3] and M. Duffy[4]

**Ilaria Cinelli** [1] **(Corresponding Author)**
National University of Ireland, Galway, Rep. of Ireland
Discipline of Biomedical Engineering
College of Engineering and Informatics
University Road
H91 TK33, Galway, Rep. of Ireland
E-mail: i.cinelli1@nuigalway.ie; i_cinelli@yahoo.it
Phone(s): +39 3492182632; +353 860 891452
ORCID: 0000-0003-1302-1188

**Michel Destrade**[2]
National University of Ireland, Galway, Rep. of Ireland
School of Mathematics, Statistics and Applied Mathematics
College of Science
H91 TK33, Galway, Rep. of Ireland
E-mail : michel.destrade@nuigalway.ie
ORCID: 0000-0002-6266-1221

**Peter McHugh**[1]
National University of Ireland, Galway, Rep. of Ireland
Discipline of Biomedical Engineering
College of Engineering and Informatics
University Road
H91 TK33, Galway, Rep. of Ireland
E-mail: peter.mchugh@nuigalway.ie
ORCID: 0000-0001-9147-3652

**Antonia Trotta**[3]
School of Mechanical & Materials Engineering
Mechanical & Materials Engineering
University College Dublin, Belfield, Dublin 4, Ireland
E-mail: antonia.trotta@ucdconnect.ie

**Michael Gilchrist**[3]
School of Mechanical & Materials Engineering
Mechanical & Materials Engineering
University College Dublin, Belfield, Dublin 4, Ireland
E-mail: michael.gilchrist@ucd.ie
ORCID: 0000-0003-1765-429X

**Maeve Duffy**[4]
National University of Ireland, Galway, Rep. of Ireland
Power Electronics Research Centre
Discipline of Electrical and Electronic Engineering
College of Engineering and Informatics
University Road
H91 TK33, Galway, Rep. of Ireland
E-mail: maeve.duffy@nuigalway.ie
ORCID: 0000-0002-1379-3279

**ABSTRACT**

*Objective:* **To investigate mechanical and functional failure of diffuse axonal injury (DAI) in nerve bundles following frontal head impacts, by finite element simulations.** *Methods:* **Anatomical changes following traumatic brain injury are simulated at the macroscale by using a 3D head model. Frontal head impacts at speeds of $2.5 - 7.5\ m/s$ induce mild to moderate DAI in the white matter of the brain. Investigation of the changes in induced-electro-mechanical responses at the cellular level is carried out in two scaled nerve bundle models, one with myelinated nerve fibres, the other with unmyelinated nerve fibres. DAI occurrence is simulated by using a real-time fully coupled electro-mechanical framework, which combines a modulated threshold for spiking activation and independent alteration of the electrical properties for each 3-layer fibre in the nerve bundle models. The magnitudes of simulated strains in the white matter of the brain model are used to determine the displacement boundary conditions in elongation simulations using the 3D nerve bundle models.** *Results:* **At high impact speed, mechanical failure occurs at lower strain values in large unmyelinated bundles than in myelinated bundles or small unmyelinated bundles; signal propagation continues in large myelinated bundles during and after loading, although there is a large shift in baseline voltage during loading; a linear relationship is observed between the generated plastic strain in the nerve bundle models and the impact speed and nominal strains of the head model.** *Conclusion:* **The myelin layer protects the fibre from mechanical damage, preserving its functionalities**.

# 1. INTRODUCTION

Current research interests in brain modelling aim at understanding head injury (Dixit 2017; Garcia-Gonzalez 2018; Samaka 2013; Horgan 2004, 2003; Wright 2012; Young 2015), axonal injury (Cinelli 2017a, 2017b, 2017c; Garcia-Grajales 2015; Jérusalem 2014; Mohammadipour 2017; Wright 2012), sport concussions (McCrory 2017), neuronal morphology (Abdellah 2018; Kanari 2018), and brain connectivity (Wazen 2014) when medical conditions are present, such as Alzheimer's disease, depression and epilepsy (Wazen 2014). Brain models vary according to application (Dixit 2017; Samaka 2013), and can be based on computed tomography and magnetic resonance tomography images (Horgan 2004, 2003) or can be based on magnetic resonance imaging (Garcia-Gonzalez 2017, 2018; Wazen 2014). Accuracy and precision of finite element models of the brain are achieved by design, where the brain anatomy is replicated by the inclusion of a certain number of layers, and where material properties aim to conform to reality (Dixit 2017; Samaka 2013). However, most macro-scale brain models do not account for a detailed representation of the micro-scale structure of nervous cells to limit the computational cost (Dixit 2017; Samaka 2013; Mohammadipour 2017; Wright 2012), although cell models have been developed in this regard (Abdellah 2018; Cinelli 2017a, 2017b; Garcia-Grajales 2015; Jérusalem 2014; Kanari 2018; Mohammadipour 2017; Wright 2012). Recent published works of finite element models of nervous cells tend to simulate both the mechanical structure and the functionality of the cell (Cinelli 2017a, 2017b; Garcia-Grajales 2015; Jérusalem 2014; Mohammadipour 2017), as its relevance has been demonstrated in experimental works (Galbraith 1993; El Hady 2015; Mosgaard 2015; Mueller 2014; Zhang 2001).

Traumatic Brain Injury is a common result of head impact. TBI is a major public-health problem generated by falls, vehicle accidents, sport injuries, military incidents, etc. (Ma 2016; Wright 2012; Zhang 2014). In Europe, fall-related, work-related and all injury-related deaths due to TBI are as high as 47.4% (Li 2016), 8.5% (Li 2016) and 37% (Majdan 2016), respectively. Brain injuries are associated with increased mortality and decreased life expectancy compared to the general population (Majdan 2016). Furthermore, people with TBI incur substantial direct (health care) and indirect (loss of productivity and care-giver related) costs (Majdan 2016). Multiple factors such as individual anatomy, head acceleration, magnitude and direction of forces, protective equipment, etc., explain the high heterogeneity of TBI (Hemphill 2015; Siedler 2014). Neurochemical, metabolic, neuroinflammation, blood perfusion and other molecular-based processes change the mechanobiology and the cellular microenvironment of the brain as a consequence of the localization of stresses following TBI (Hemphill 2015; Kan 2012). The challenge in understanding the biomechanics of TBI leads to an increase in difficulty in treating and preventing the development of cognitive and behavioural problems (Hemphill 2015; Kan 2012).

Additionally, TBI has effects at the *cellular level*. The neuropathology of TBI includes focal damage of brain tissue or widespread axonal injury (Jérusalem 2014; Wright 2012). TBI-induced dynamic deformations increase the risk of axonal stretch and shear injuries to axons scattered throughout the brain parenchyma (Petrosino 2009; Wright 2012), generating structural and functional damage (such as leaking nerve membranes (Yu 2012) and cytoskeleton disruption (Hemphill 2015; Smith 1999; Tang-schomer 2017)), leading to rupture of the axon (Hemphill 2015; Siedler 2014; Smith 1999). Strain and strain-rate are known to play important roles in the induced-electrophysiological impairments and functional deficits at the axonal level (Boucher 2012; Geddes 2003; Jérusalem 2014).

In particular, mild and severe TBI impacts lead to *Diffuse Axonal Injury* (DAI), which refers to the damage experienced by neural axons in the deep white matter regions of the brain (Ma 2016; Wright 2012). DAI is associated with a high risk of developing future neurodegenerative disease (Hemphill 2015; Kan 2012), and the progressive course of DAI is responsible for long-lasting neurological impairments associated with high rates of mortality (Lajtha 2009; Smith 2000; Wang 2010). Currently, no clinical treatments and prognosis can be used against DAI because of the complexity in diagnosis when using medical imaging, due to haemorrhages, hematomas and tissue lesions of the neighbouring injured area (Hemphill 2015; Lajtha 2009; Ma 2016; Wright 2012). In effect, DAI can only be established post-mortem.

Computational models of TBI biomechanics can simulate brain trauma based on the principles of mechanics. By replicating head impact dynamics, they enable a detailed investigation of the mechanical and physiological changes linked to anatomical and functional damage of the brain. Modelling presents itself as a tool that could aid in diagnosis as it allows for the evaluation of mechanical and physiological quantities in the brain tissue that cannot be detected by current medical technology.

With the purpose of enhancing the understanding of electro-mechanical DAI occurrences, in this work we adopt a multi-scale approach, and use two independent models to replicate the anatomical and functional changes induced by TBI events: (i) at the *macroscale*, the induced anatomical changes are simulated by using an advanced 3D biomechanical Head Model (the University College Dublin brain trauma model – UCDBTM (Horgan 2004, 2003)); (ii) at the *microscale*, the structural and functional changes of complex electro-mechanical impairments at the cellular level are simulated by using a 3D coupled electromechanical Nerve Bundle Model (Cinelli 2017a, 2017b). These models replicate the brain macro-environment and the neural micro-environment, respectively. The sequential use of two independent models is needed to limit the computational cost that would arise from the combined modelling of macro and micro features within the same 3D model.

In this paper, we simulate short-term frontal head impacts by using the Head Model to estimate the deformation of the white matter regions at the instant of impact. Then, the magnitudes of TBI-induced strains in the Head Model are used to determine the displacement boundary conditions to be applied to the Nerve Bundle Model. The worst case condition of uniaxial stretching is applied to the fibre bundles, where it has been found that tensile axonal strain is the most realistic mechanism for generating DAI (Bain 2000; Jérusalem 2014; Wright 2012). The variation of the membrane voltage (in terms of its peak and baseline values) is investigated during and after the applied elongation. In this way, electrophysiological and structural occurrences are simulated at the axonal level using realistic TBI elongation simulations. DAI-induced neural strains and voltage changes are analysed using the Nerve Bundle Model in relation to strain levels predicted by the Head Model for different impact speeds.

In the field of head trauma biomechanics, the Head Model developed at University College Dublin (Horgan 2003) is a 3D finite element (FE) representation of the human head complex, proposed as a tool for the assessment of brain injury mechanisms (Horgan 2004, 2003). The main anatomical features of the UCD Head Model include the cerebrum, cerebellum and brainstem, intracranial membranes, pia, cerebrospinal fluid layer, dura, a varying thickness three-layered skull (cortical and trabecular bone layers), scalp and the facial bone (Horgan 2003), see Fig. 1. A parametric analysis, undertaken using Abaqus 5.8, was performed to investigate the influence of different mesh densities on the model geometries and material properties (Horgan 2003). The Head Model consists of a total of 26,913 nodes and 28,287 elements, including linear quadrilateral, hexahedral and triangular elements, see Fig. 1. It has been validated against a series of cadaveric head impact experiments, simulating the Nahum's test (Horgan 2003), and Trosseille's and Hardy et al.'s tests, see account by Horgan et al. (Horgan 2004). The validation includes both rotational and translational acceleration components (Horgan 2004, 2003). In the current study, we use the Head Model to simulate frontal head impacts only, where rotations and accelerations are neglected.

Recent experimental evidence of neural activity highlights *complex electro-mechanical phenomena* happening at the nerve membrane layer (Alvarez 1978; Cinelli 2017a, 2017; Galbraith 1993; Geddes 2003; Mueller 2014; Zhang 2001) during signalling (Mosgaard 2015; Zhang 2001). The inclusion of an accurate representation of DAI-related electrophysiological impairments is needed to improve diagnosis, treatment and prognosis of related pathologies (Jérusalem 2014; Lajtha 2009; Ma 2016; Wright 2012). Our Nerve Bundle Model is a 3D idealized representation of a nerve bundle (Cinelli 2017a, 2017b, 2017c), located in the deep white matter of the brain (Wright 2012), see Fig. 2. The 3D bundle is made of four identical aligned axons whose diameter is within the range of small fibres of the human corpus callosum only (Björnholm 2017), as discussed in (Cinelli 2017a), see Fig. 2. Each fibre is made of extracellular media (ECM), see Fig. 2 (c), intracellular media (ICM), see Fig. 2 (e), and a membrane, see Fig. 2 (d). At the nerve membrane layer, the Nerve Bundle Model includes a fully coupled 3D electro-mechanical representation of the neural activity, combining piezoelectricity and electrostriction with changes in strain, including total strain (elastic, electro-thermal equivalent and plastic strain) (Cinelli 2017a, 2017b). The piezoelectric effect corresponds to a linear variation of the electrical polarization of a medium linearly with the applied mechanical stress, as seen in nervous cells (Mosgaard 2015; Mueller 2014; Zhang 2001), while electrostriction is a quadratic effect (Alvarez 1978; Mosgaard 2015; Mueller 2014; Zhang 2001) which refers to the displacement of a dielectric under an applied electric field.

We consider the case of two scaled nerve bundle models with a ratio of 2:1, where the nerve fibres inside have the same ratio, while the thickness of the nerve membrane is maintained constant. The neurite radii of the small bundle (SB) model are: $a_{ICM} = 0.477\ \mu m$, $a_M = 0.480\ \mu m$ and $a_{ECM} = 0.500\ \mu m$ (Cinelli 2015, 2017c). The neurite radii of the second, bigger bundle (BB) are double those of SB, while the membrane thickness is the same ($3nm$). Here, only the cases of a fully unmyelinated bundle or a fully myelinated bundle are considered. In the case of myelinated fibres, the nerve membrane section, see Fig. 2 (d), is periodically-partitioned along the fibre length, similar to the histologic section of a myelinated fibre, see Fig. 1 (d.1) and (d.2) (Cinelli 2017a, 2017b). The width of the piecewise conductive membrane regions (or Ranvier's nodes) is $0.002\ \mu m$ and the internode distance is $1\ \mu m$ (Cinelli 2017; Einziger 2003), see Fig. 1 (d.2). Different electrical properties are assigned to the regions of myelin and Ranvier's node, respectively (Einziger 2003).

This work is a development on the work reported in (Cinelli 2017a, 2017b, 2017c, 2017d). In contrast to (Cinelli 2017a, 2017b, 2017c, 2017d), the boundary conditions used in the Nerve Bundle Model are directly linked to frontal head impacts simulated with the Head Model. Thus, control over the boundary conditions of both models aims at simulating realistic clinical events, speeding up the transfer of the findings from computational studies to clinical care.

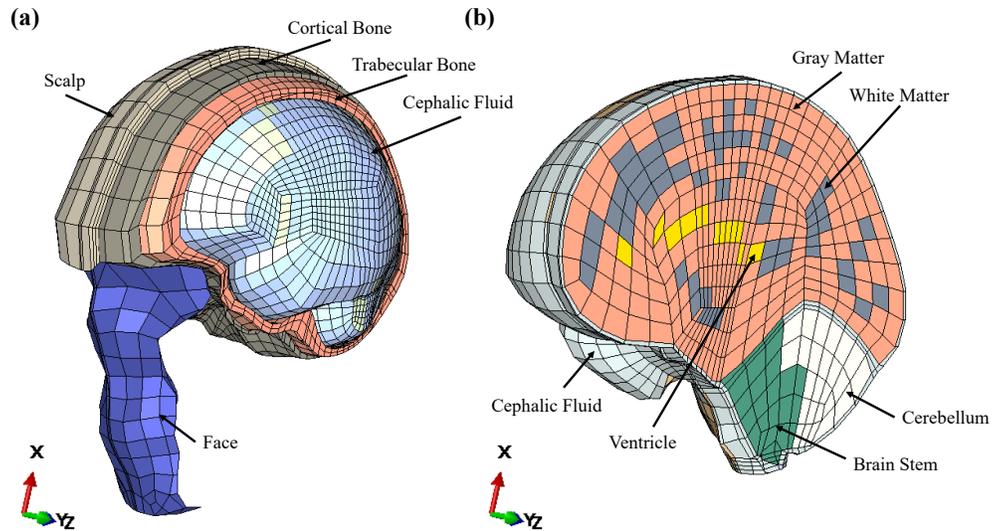

**Fig. 1** The UCD Head Model (Horgan 2004, 2003): in (a), the face, scalp, cortical and trabecular bone, and cephalic fluid; in (b), grey matter, white matter, cerebellum and brain stem.

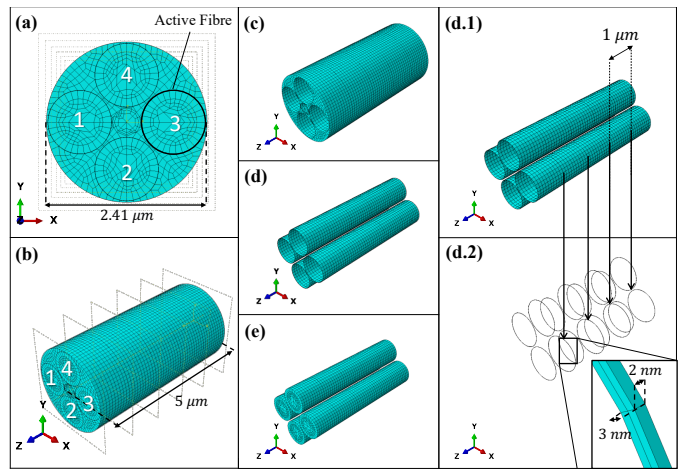

**Fig. 2** (a): frontal view and (b): isometric view of the 3-layer nerve bundle made of four fibres. Fibre #3 is the active fibre, i.e. the fibre activated by a Gaussian voltage distribution (El Hady 2015). Fibres #1, #2 and #4 are activated by the charges diffusing from Fibre #3. (c): the ECM; (d): the membrane; (e): the ICM. In the case of myelinated fibres, the membrane layer is periodically partitioned along the fibre length to model the insulation sheath of the myelin layer, see (d.1), and the Ranvier node, see (d.2). The myelin layer length is 1 μm and the Ranvier node length is 2 nm, while the radial thickness of the layer is equal to 3 nm (Cinelli 2017, 2015; Einziger 2003) (figure reproduced from (Cinelli 2017)).

## 2. METHOD

### 2.1. Boundary Conditions

We simulate a frontal impact to induce axonal injuries in the deep white matter of the brain. The Head Model is launched freely with an initial velocity (from 2.5 to 7.5 $m/s$) against an encastré plane (i.e. the floor) (Horgan 2003). The speed values are within the range of values considered to induce mild and moderate DAI in white matter (Wright 2012). We assume that the loading axis of the velocity impact is aligned parallel to the force of gravity. Since the neck is not included in the model, a free boundary condition is used to simulate a frontal head impact and only short-duration impact responses ($<$ 6 $ms$ (Horgan 2003)) are considered (Horgan 2003).

For the Nerve Bundle Model, the encastré boundary condition is applied at the origin of the model at one end, and no rotations are allowed, while a displacement boundary condition is applied at the opposite end (as in uniaxial elongation). Then, an upper-threshold stimulation voltage with a Gaussian distribution is applied on Fibre #3 along its length, see Fig. 2, while the other fibres are activated only if the diffused charges from Fibre #3 generate an input voltage higher than the modulated threshold (Cinelli 2017b, 2017c). The 3D distribution of charges on Fibre #3 modulates the activation of the other fibres, see Fig. 2.

Invoking the *macro-micro* link, the magnitude of the displacement boundary conditions of the Nerve Bundle Model are taken from the nominal strain value found in the white matter regions of the Head Model following impact. Frequency-independent loading conditions are considered throughout, after an initial steady-state interval (lasting about $2ms$). The mechanical loads are applied to the Nerve Bundle Model from 2 $ms$ to 67 $ms$, as instantaneous loading conditions, and the model is set to run for 140 $ms$ so that the effects of plasticity can be observed post-loading.

### 2.2. Material Properties

Details of the Head Model formulation and material properties can be found in the papers by Horgan et al. (Horgan 2004, 2003), but to summarise, the model utilises linear viscoelasticity combined with hyperelasticity and large deformation kinematics to represent the brain tissue.

For the Nerve Bundle Model we assume incompressible rate-independent isotropic mechanical behaviour (El Hady 2015), as described in Cinelli et al. (Cinelli 2017a, 2017b). Further, we include plasticity, by assuming the same isotropic plastic behaviour for the nerve membrane, ICM, and myelin layer. The yield stress is calculated with an engineering strain equal to 21 % (Bain 2000) and a Young's Modulus equal to $1GPa$ (El Hady 2015). Strain hardening is assumed to occur up to a strain of 65 % (Smith 1999). Thus, the engineering strain and engineering stress values are $(0.21, 0.21\ GPa)$ and $(0.65, 0.65\ GPa)$ for the yield strain limit and strain hardening, respectively. Beyond 65 % strain, the stresses are assumed to be constant.

The electrical model parameters for unmyelinated and myelinated fibres are taken from (Cinelli 2015, 2017c) and (Jérusalem 2014), respectively. The piezoelectric effect is only relevant in the through-thickness direction, represented here with orthotropic piezoelectric constants of approximately 1 nm per 100 mV (Zhang 2001) in the thickness direction and zero in the longitudinal and circumferential directions, while the electrical capacitance per unit area changes as the square of the voltage (Cinelli 2017a, 2017b; El Hady 2015).

### 2.3. Implementation

The Head Model is implemented as a dynamic analysis in Abaqus CAE (Horgan 2004, 2003) (using Abaqus/Explicit), to allow for the representation of head impact dynamics and the associated deformation of the brain tissue due to impact.

The Nerve Bundle Model is implemented as a quasi-static analysis in Abaqus CAE 6.13-3, as described in Cinelli et al. (Cinelli 2017a, 2017), using Abaqus/Standard. The implementation of the coupled Hodgkin and Huxley (HH) model is shown in Fig. 2 (on the right), and contrasted to the original, uncoupled HH model (on the left) (Cinelli 2017a, 2017b). By using the electro-thermal equivalence (Cinelli 2017a, 2017b, 2017c), implementation of the coupling between neural activity and the generated strain can be achieved in 3D. The equivalent electrical properties change independently at the nerve membrane of each fibre, based on the spike initiation, strain and voltage generated at the same membrane (Cinelli 2017a, 2017b). In the coupled model (Cinelli 2017b), the membrane neural activity changes in response to the membrane voltage and total strain (elastic, electro-thermal equivalent and plastic),$\varepsilon$ at the membrane (Hodgkin 1952; Jérusalem 2014), while the electrical capacitance per unit area,$C_m$, changes with the square of the voltage (Cinelli 2017a, 2017b, 2017c), see Fig. 2. The HH resting voltage potentials of sodium, $E_K$, and potassium, $E_{Na}$, change due to voltage and strain at the nerve membrane (Boucher 2012; Jérusalem 2014), and hence the threshold of spike initiation changes as prescribed by Hodgkin and Huxley (Hodgkin 1952). The reversal potential of the leak ions $E_{l^-}$ is not influenced by the strain but varies based on changes in the gradient concentrations of potassium and sodium across the membrane (Jérusalem 2014). The changes in ion conductance for sodium, $G_K$, and potassium, $G_{Na}$, follow the changes in the respective reversal potentials, as in (Hodgkin 1952).

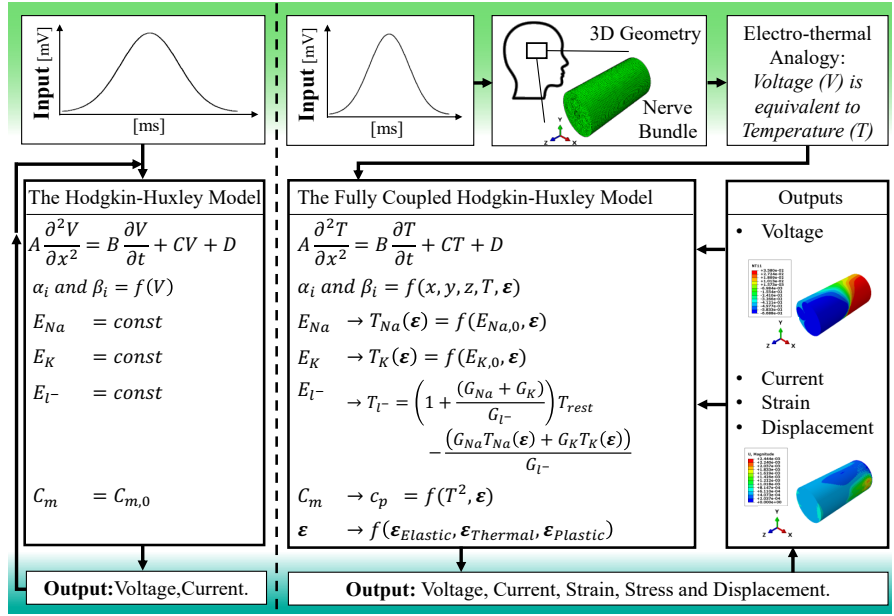

**Fig. 3** Flowchart of the code describing the active behaviour of the nerve's membrane: on the left, the HH dynamics (Hodgkin 1952) and on the right, the fully coupled HH dynamics. A, B, C and D are general numerical values of specific physical quantities, used to established the electro-thermal equivalences upon which this modelling approach is based (Cinelli 2017b). Here, a Gaussian voltage distribution elicits the action potential in a 3D model of a nervous cell. By using electro-thermal equivalences, the HH model is implemented as an equivalent thermal process, in which the membrane's conductivity changes as in (Hodgkin 1952) and the capacitance, $C_m$, changes as in (Cinelli 2017a). The HH parameters are changing based on the temperature and strain at the membrane (Cinelli 2017a, 2017 b, 2017v). The strain $\varepsilon$ generated in the model is a function of temperature, T, and thermal expansion coefficients (Cinelli 2017a, 2017b, 2017c). Voltage, current, strain and stresses distribution are only a few of the 3D results released by Abaqus by equivalence. (Figure reprinted from (Cinelli 2017b)).

Table I shows the parameters used in this model for the steady state and time varying conditions, and their corresponding value, taken from published experimental literature. Our model accounts for the HH dynamics (Hodgkin 1952), implemented by using the thermal analogy of the neural electrical activity in finite element analysis (Cinelli 2017a, 2017b, 2017c). Specific parameters of the ionic reversal potentials vary with the applied strain in time varying conditions, as described by Jerusalem A. et al. 2014 (Jérusalem 2014). Then, due to the variation in thickness (Galbraith 1993; El Hady 2015; Mosgaard 2015; Mueller 2014; Zhang 2001), the membrane capacitance per unit area of a nerve fibre varies as the square of the membrane voltage, $V$, (Alvarez 1978). More details about the simulation process and formulas, upon which the electro-mechanical coupling is established, are shown in (Cinelli 2017a, 2017b, 2017c).

Depending on the applied boundary conditions, the model is able to simultaneously generate data about the displacement, strain, voltage (equivalent to temperature (Cinelli 2017c)), and stresses (Cinelli 2017a, 2017b). Additional quantities can be also released depending on the application. For the purposes of this paper, the strains are applied at the nerve bundle, see Boundary Condition section, while the voltage is the unknown variable. In previous work, voltage boundary conditions have been used to quantify strain, displacement and stresses (Cinelli 2017c).

Table I: Parameters of the fully coupled HH dynamics.

| Parameters | Symbol | Unit | Formula | Value | Ref. |
|---|---|---|---|---|---|
| **Steady state** | | | | | |
| Ionic reversal potentials | | mV | | | |
| *Sodium* | $E_{Na,0}$ | | | -115 | (Hodgkin 1952) |
| *Potassium* | $E_{K,0}$ | | | 12 | (Hodgkin 1952) |
| *Leak Ions* | $E_l, 0$ | | | -10.63 | (Hodgkin 1952) |
| Ionic conductance | | $mS\ cm^{-2}$ | | | |
| *Sodium* | $\bar{g}_{Na}$ | | | 120 | (Hodgkin 1952) |
| *Potassium* | $\bar{g}_K$ | | | 36 | (Hodgkin 1952) |
| *Leak Ions* | $\bar{g}_l$ | | | 0.3 | (Hodgkin 1952) |
| Membrane capacitance per unit area | $C_m(0)$ | $\mu F/cm^2$ | | 1 | (Hodgkin 1952) |
| **Time varying** | | | | | |
| Ionic reversal potentials | | | | | |
| *Sodium* | $E_{Na}(\varepsilon_m)$ | | $E_{Na}(\varepsilon_m) = E_{Na0}(1-(\varepsilon_m/\tilde{\varepsilon})^\gamma)$ | | (Jérusalem 2014) |
| *Potassium* | $E_K(\varepsilon_m)$ | | $E_K(\varepsilon_m) = E_{K0}(1-(\varepsilon_m/\tilde{\varepsilon})^\gamma)$ | | (Jérusalem 2014) |
| Membrane capacitance per unit area | | $\mu F/cm^2$ | $C_m(V) = C_m(0)[1+\vartheta(V+\Delta V)^2]$ | | (Alvarez 1978) |
| Strain threshold | $\tilde{\varepsilon}$ | % | | 21 | (Bain 2000; Jérusalem 2014) |
| Exponential value | $\gamma$ | | | 2 | (Jérusalem 2014) |
| Fractional increase in capacitance per square volt | $\vartheta$ | $V^{-2}$ | | 0.036 | (Alvarez 1978) |

# 3. RESULTS

Following a frontal head impact with initial speed $v$ $[m/s]$, the maximum principal value of the nominal strains ($NE$) [%], found in the white matter region of the Head Model, vary with $v$, see Fig. 4. Although the inclusion of large deformation kinematics in the head model, the $NE$ variation is near linear. This justifies the investigation of the electrophysiological features of the Nerve Bundle Models based on the quasi-static assumption where only small deformations are considered.

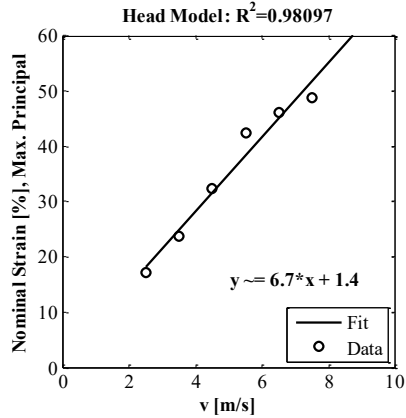

**Fig. 4** Variation in the maximum principal value of the nominal strains [%] in the Head Model vs. the speed impact values, v [m/s], used in the frontal head impacts of the Head Model. Linear regression fit also included.

Fig. 5 shows the Head Model, in (a)-(c), and the Nerve Bundle Model, in (d)-(i). This figure refers to the case of impact at 7.5 $m/s$, only. The head impacts the floor at 7.5 $m/s$ as shown in (a), where the gravity force is aligned to the $z$ −axis of the Head Model. At the time of contact with the floor (i.e. 3.75 $ms$), (b) and (c) show the cerebellum and white matter region of the model, respectively. The contour plots refer to the $NE$ in the Head Model. From this, the NE in the Head Model is used to generate the magnitude of the displacement boundary conditions of the Nerve Bundle Models. Figs. 5 (h) and (i) show the differences in the mesh and size of the model use for the SB and BB cases in this work. The voltage distribution ($NT11$), equivalent to temperature (Cinelli 2017a, 2017b, 2017c), is shown in (d) for BBUN, in (e) for BBMY, in (f) for SBUN and in (g) for SBMY. Here, the voltage refers to the maximum value reached during elongation simulation, in which 48.7 % strain determines the magnitude of the displacement boundary condition, see (c), and refers to the case of 7.5 $m/s$ impact, see (a).

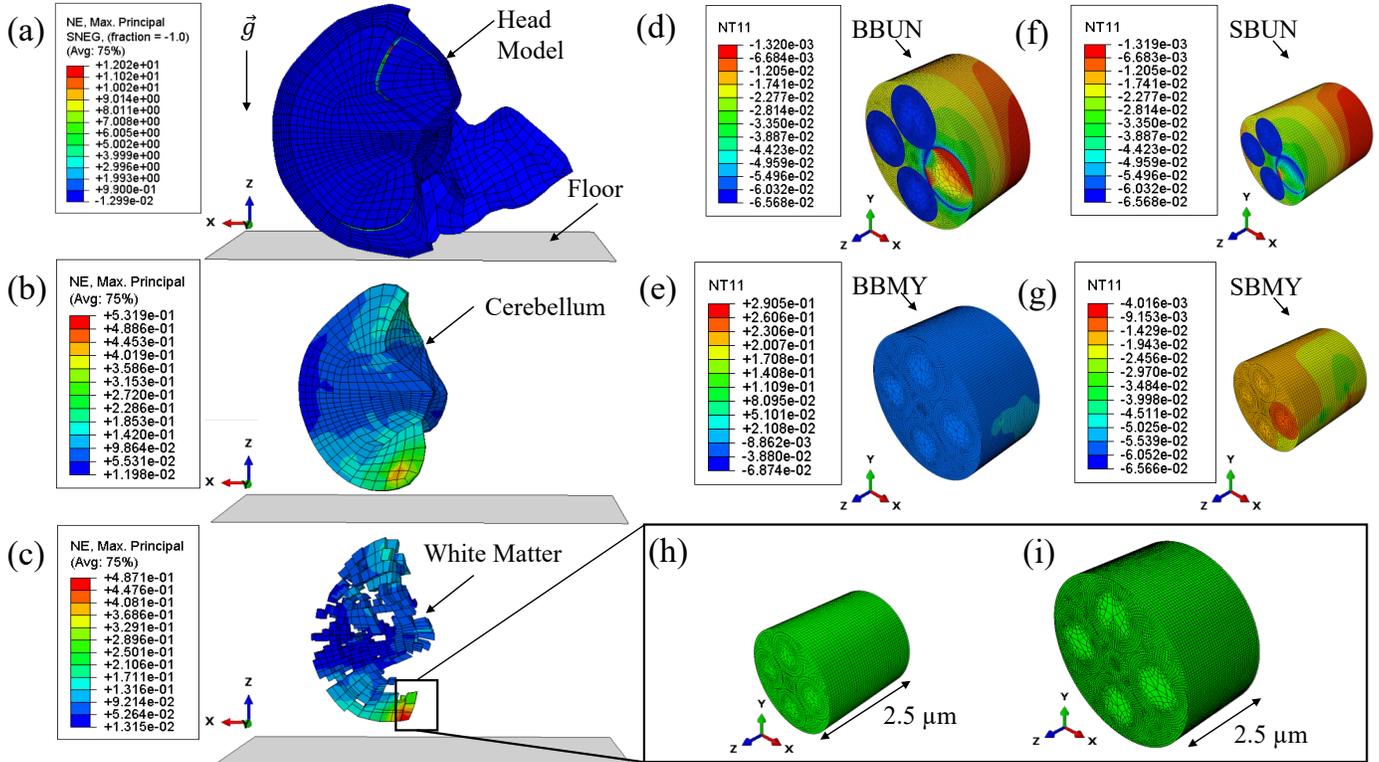

**Fig. 5** (a)-(c) the Head Model (Horgan 2004, 2003) and in (d)-(i) the Nerve Bundle Model. The head impacts the floor at 7.5 m/s, in (a), and, at the instant of contact with the floor, maximum principal nominal strains are shown for the cerebellum, see (b), and white matter regions of the Head Model, see (c). A displacement boundary condition is applied to the Nerve Bundle Model with magnitude determined from the maximum value of the strain found in the white matter, see (c), for the same case of impact. The voltage (NT11) is shown during elongation, at the peak of the membrane potential. The voltage distribution is shown for: BBUN (d), BBMY (e), SBUN (f) and SBMY (g). In (h) and (i) a representation of the mesh and geometry of SB and BB used in this work. The maximum value of each contour plot is reported here in brackets for clarity. For the Head Model the maximum of the nominal strain is 12.02, for the cerebellum is 0.5319, and for the White Matter is 48.71. Then, peak of the voltage in the BBUN is -1.320 mV, for the BBMY is 0.2905 mV, for the SBUN -1.309 mV and for the SBMY is -4.016 mV.

Then, the alteration of mechanical and electrical variables of the Nerve Bundle Models is shown in Figs. 6-7. The cases of small unmyelinated (SBUN), small myelinated (SBMY), big unmyelinated (BBUN) and big myelinated (BBMY) bundles are considered. A head impact at 2.5 $m/s$ generates a mild-intensity axonal injury at the nerve axon level, because the magnitude of the $NE$ is lower than the threshold considered, 21% (Bain 2000), for initiating plasticity, and equal to 17.1 % (see Fig. 4). In contrast, impacts at 3.5 to 7.5 $m/s$ generate strains within 23.7 % to 48.7 % (beyond the yield strain limit), inducing moderate-intensity axonal injury at the nerve axon level (Bain 2000; Smith 1999). The plastic strain values of the Nerve Bundle Model are permanent strains generated during elongation, and consequently are found to remain after loading (not shown here). Fig. 6 (a)-(b) and (d) shows that for SBUN, SBMY and BBMY, the magnitude of the maximum principal values of the plastic strain ($PE$) (%) read at the nerve membrane, have a near-linear relation with the $NE$ found in the Head Model, and that the $PEs$ vary between 0 and 15%. In contrast, in the BBUN, the $PEs$ are found to be within 0% and 150%, showing a more non-linear relation with the $NE$, see Fig. 6 (c). It is found that bigger bundles undergo larger deformations compared to small bundles, and that the myelin sheath has an important role in redistributing the applied strains, preserving the mechanical structure of the fibre (Cinelli 2017a, 2017b). In the BBUN, the entire nerve membrane layer is exposed to the applied deformation during elongation, while, in the BBMY, the myelin sheath protects the Ranvier's node regions from higher deformation by holding up part of the applied strains. Thus, the $PEs$ read at the Ranvier's node regions in BBMY are lower than the $PEs$ found in BBUN.

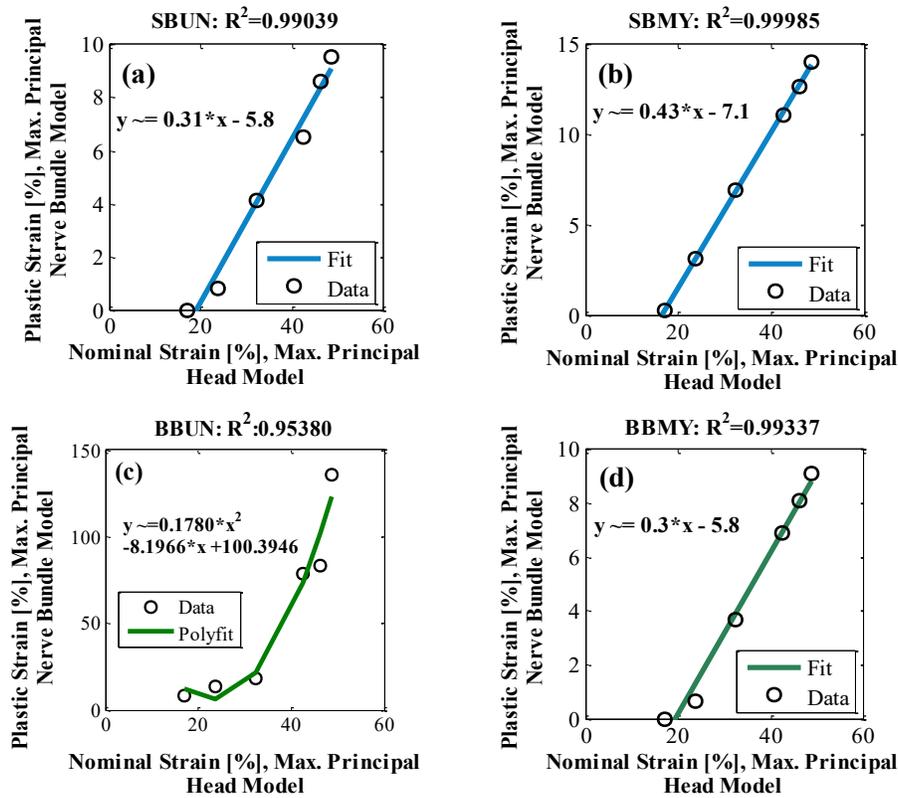

**Fig. 6** Variation in the maximum principal value of the plastic strains [%] in the Nerve Bundle Model, read at the nerve membrane, vs. the maximum principal value of the nominal strains [%] in the Head Model for (a) SBUN, (b) SBMY, (c) BBUN and (d) BBMY. Regression fits also included.

Fig. 7 shows the membrane potential peak and baseline values on Fibre #3 of the Nerve Bundle Models (a)-(c) during loading, and (b)-(d) after loading; values are plotted against the maximum principal value of the *NE* found in the white matter region of the Head Model, following impact. In Fig. 7 (a)-(c), each point is the peak and baseline voltage respectively, read at the nerve membrane when the elongation displacement boundary condition is applied on one end of the Nerve Bundle Model. The values are taken at the position where maximum displacement along the fibre middle axis occurs.

The membrane voltage (peak and baseline) changes in relation to the total strains read at the membrane, along the fibre length (Cinelli 2017b) and from the corresponding changes in the ionic reversal potentials (Cinelli 2017a, 2017b), see Fig. 3. For total strain lower than the 21% (Cinelli 2017a,) the ionic reversal potentials vary to produce a different homeostatic ionic gradient across the nerve membrane. For higher strain, the reversal potentials reach a saturation level, simulating loss of charges across the nerve membrane due to permanent deformations at the nerve membrane. This is seen in terms of saturation in membrane voltage levels with increasing *NE* for all cases except SBUN in Fig. 7(a)-(c).

The voltage differences found during and after elongation, see Fig. 7, arise from the elastic recovery that occurs as soon as the load is removed, while the *PEs* remain after loading. Results found after-loading, shown in (b)-(d), are taken at the same node as during loading for ease of comparison. The dashed line refers to the ability of the fibre to generate signals (here, it is found for BBMY only), in contrast with small oscillations of the membrane voltage around the baseline value (solid lines). The levels of oscillation can be deduced by comparing corresponding peak and baseline values, where only the BBMY case shows an appreciable difference both during and after loading, see Fig. 7. Again, the reason is because the myelin sheath layer redistributes the plastic strain around the Ranvier's node regions, thus the corresponding changes in the ionic reversal potentials are lower than the other cases. In all cases except the BBUN, there is a general increase in membrane voltage with *NE* after loading, while the non-linear variation for BBUN is explained by the non-linearity in *PE* (see Fig. 6).

To further illustrate how these results relate to the impact conditions, the supplementary material shows results of *PE* in the Nerve Bundle Model vs. the speed of impact of the Head Model, and the membrane voltage peak and baseline vs. the speed of impact.

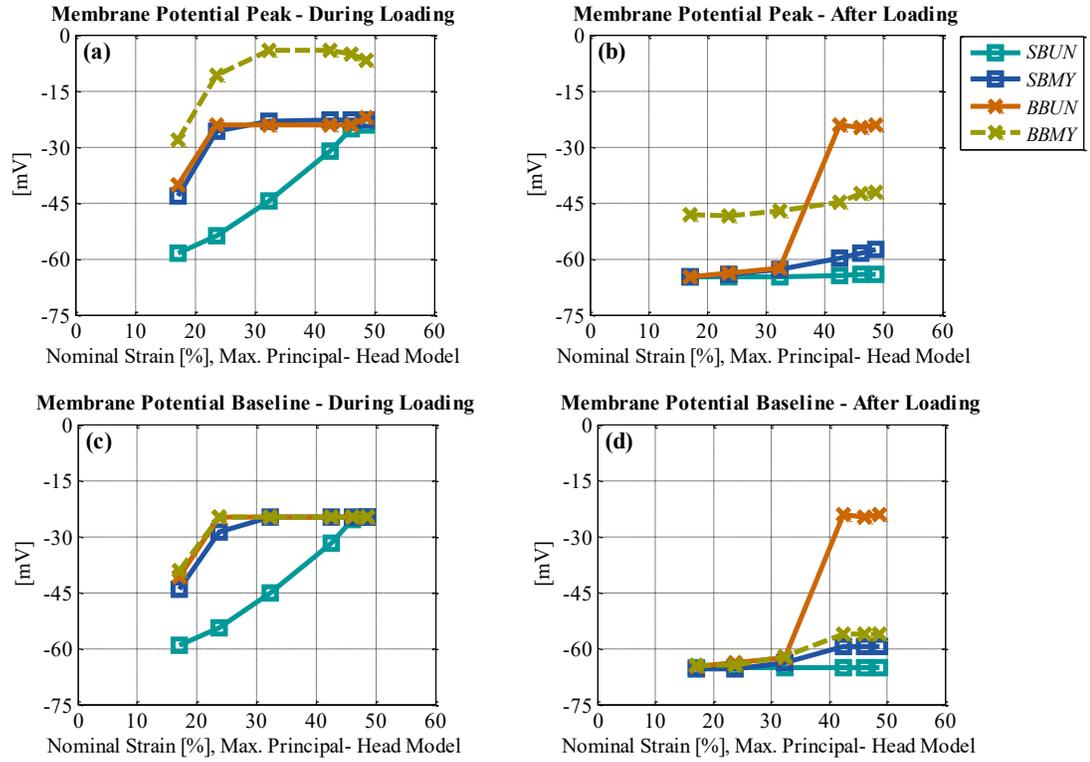

**Fig. 7** The membrane potential peak [mv], read at the nerve membrane, vs. the maximum principal value of the nominal strains [%] found in the Head Model; (a) and (c) show the potential values during elongation, while (b) and (d) show the potential values after elongation. On the top, (a) and (b), are for the membrane potential peak, read on Fibre#3. On the bottom, (c) and (d), are for the membrane baseline.

# 4. DISCUSSION

DAI arises from damage of the white matter following TBI (Ma 2016; Wright 2012). The current literature on the subject discusses at length the neurological and non-neurological consequences of DAI (Kan 2012; Lajtha 2009; Smith 2000; Wang 2010), the need for need for better diagnosis (Hemphill 2015; Lajtha 2009; Ma 2016; Wright 2012), and the lack of suitable treatments (Hemphill 2015; Lajtha 2009; Wright 2012).

Within this context, computational modelling can be used to simulate complex anatomical and functional damage induced by TBI at different scales, with the goal of improving the understanding of brain injuries and the quality of clinical care. In this work, induced damage following TBI is simulated at the macroscale by using a Head Model (Horgan 2004, 2003), while damage to cellular mechanisms, i.e. the electro-mechanical impairments of diffuse axonal injury (DAI), is simulated by using a Nerve Bundle Model (Cinelli 2017a, 2017b, 2017c, 2017d). The use of two independent, although linked, models is needed to have full control over the applied boundary conditions at the different size scales. Investigation of the structural damage caused by strain levels determined during impact at the macroscale is implemented using an estimation of the plastic strain at the nerve membrane, which is the irrecoverable component of the total strain. The corresponding voltage changes simulated in the Nerve Bundle Model are a measure of the functional damage at the axonal level.

The results show that plastic strains found in the Nerve Bundle Model are linearly related to both the nominal strains generated in the white matter of the Head Model during impact, and the impact speed values themselves, see Fig. 4 and 6. Although large deformation is assumed in the Head Model, the quasi-static assumption of the Nerve Bundle Model is justified by the linear deformation of the head regions, following impact, see Fig. 4.

By elongating the bundles, it is interesting to note that permanent strains at the membrane are higher in the large unmyelinated bundle, while they are lower than 21% in the other Nerve Bundle Models, see Fig.6. Following a high speed frontal head impact, mechanical failure may occur in larger unmyelinated nerve bundles, due to the high plastic strain produced at the membrane, see Figs. 6 and 7 (a)-(d). In this work, physical disconnection of fibres within the nerve bundle models (or axotomy (Wang 2011)) is not accounted in the simulation for the range of speeds considered. However, we assume that high plastic strains in the bundle are indicative of disconnection and so, mechanical failure.

Related to this, the functionality of the fibre is also affected by the (axial) component of the total strain, while it also depends on the high number of charges exchanged per unit area (Cinelli 2017a, 2017b, 2017c). The *macro-to-micro* link allows for analysing the structural and functional failure in nerve bundles with different calibre, following frontal head impacts. Thus, the permanent alterations of the membrane potential are DAI-induced electrical changes that can be linked directly both to the speeds of impact and to the nominal strain of the white matter.

Deforming the bundle, permanent deformations occur at the nerve membrane and at the Ranvier's node regions of unmyelinated and myelinated nerve fibres, respectively, which in turn change the ionic reversal potentials and consequently the homeostatic gradient of the nerve membrane. The shift of the membrane potential to positive values is proportional to the level of stretch applied to the bundle (Cinelli 2017b), where the ionic gate channels are physically stretched and kept opened by the applied loads. Due to the magnitude of nominal strains applied (beyond the yield strain limit), these structural and functional changes are not reversible after loading for all the conditions considered here. Small fluctuations of the membrane potential around the voltage baseline could be interpreted as the loss of the ability of the membrane to generate an action potential, thereby leading to functional failure. The use of the *macro-to-micro* link for analysing functional damage could be thought as an estimation of the cellular functionality following DAI, without the use of invasive devices.

For speed impact values of up to 7.5 $m/s$, it is found that after loading (i.e. without the elastic component of the total strain) the voltage alterations simulated by the fully coupled Hodgkin and Huxley model (Cinelli 2017b, 2017b) (see Fig. 3) are shown to be near-linearly related with the nominal strains in the Head Model for the small bundles (SBUN and SBMY) and big myelinated bundle (BBMY). A higher order relation is needed to describe the non-linear voltage alterations caused by strain levels beyond the yield strain for the big unmyelinated bundle (BBUN), and for all bundle types during loading.

In contrast to unmyelinated fibres, myelinated fibres preserve the ability to conduct signals even at high deformations applied to the bundle, because the mechanically induced voltage alterations are significant at the Ranvier node regions only, rather than along the whole fibre length. The myelin layer is thought to protect the fibre (Cinelli 2017a, 2017b) by constraining the deformation at the Ranvier node regions, allowing for a faster recovery of the normal membrane potential baseline value after loading. Therefore, the myelin sheath seems to protect the fibre from mechanical and functional failure. Here, the large myelinated bundle is the only bundle type still able to carry a signal after impact, although there is a significant shift in membrane voltage (and therefore change in reversal potentials) during loading.

As seen in experimental studies (Galbraith 1993; Smith 1999, 2000), strain-rate dependence in elongation is shown to play an important role in understanding damage of axonal cytoskeleton, changes in ionic gating channels and disconnection. Future work can consider further development of the current models to include strain rate dependence in elongation tests at the cellular level, in addition to rotation and acceleration in head impacts. Finally, the inclusion of a more realistic geometry of the nerve bundle and the inclusion of mechanical anisotropy in the fibre would lead to a more accurate result.

## 5. CONCLUSION

This work reports the results of an investigation of electro-mechanical impairments of DAI, following TBI events of frontal head impacts. The findings of this work could be easily transferable to clinical applications, thanks to the close link between boundary conditions in both models. The main findings can be summarized as:

- At high impact speeds (that cause high nominal strain in the white matter), disconnection of fibres is more likely to occur in large unmyelinated bundles due to the high plastic strains found at the nerve membrane;
- Signal propagation is preserved in the large myelinated bundle;
- There is a linear relation between the generated plastic strain of the nerve bundle model and the impact velocity and the nominal strains of the head model;
- The relation between the membrane voltage peak after loading and the nominal strains in the head model is near-linear in small bundles (regardless of type) and in big myelinated bundles;
- The myelin layer protects the fibre from mechanical failure, preserving its functionalities.

This model can contribute to the understanding DAI occurrences to improve diagnosis, clinical treatments and prognosis by simulating the mechanical changes accompanying the changes in signal transmission.

## LIMITATIONS

In this work, the calibre of the fibres is within the range of small fibres of the human corpus callosum only (Björnholm 2017), as discussed in (Cinelli 2017a). As shown in (Björnholm 2017; Kanari 2018), neuronal morphology varies in the nervous system. A wider range of calibre might be accounted in future studies for a more accurate representation of different brain regions and for representing the spatial distribution of mechanical and function failure at micro-scale through the brain tissue layers (Cinelli 2017a).

Then, for simplicity, the assumption of incompressible rate-independent isotropic mechanical behaviour is chosen for describing the mechanics of nerve bundles (Cinelli 2017a, 2017b, 2017c; El Hady 2015), in contrast to the different biophysical phenomena chosen for describing the electro-mechanical coupling and electrical neural activity (Cinelli 2017a, 2017a, 2017). Thus, this work is not comprehensive enough for replicating viscoelastic behaviour or mechano-sensing properties as observed to occur in nervous cells (Geddes 2003; Mary 2012). A different choice of mechanical properties and calibre might vary the results shown in this paper (Cinelli 2017a).

## ACKNOWLEDGMENT

The authors acknowledge funding from the Galway University Foundation, the Biomechanics Research Centre and the Power Electronics Research Centre, College of Engineering and Informatics, NUI Galway (Galway, Republic of Ireland).

## CONFLICT OF INTEREST STATEMENT

The authors declare that the research was conducted in the absence of any commercial or financial relationships that could be construed as a potential conflict of interest.